\documentclass{aa}  

\usepackage{hyperref}
\hypersetup{unicode=true, colorlinks=true, linkcolor=magenta, citecolor=blue, filecolor=red, urlcolor=blue}

\usepackage{ulem}

\usepackage{color}
\usepackage{array}
\usepackage{graphicx}
\usepackage{caption}
\usepackage{subcaption}
\usepackage{booktabs}
\usepackage{adjustbox}
\usepackage[toc,page]{appendix}
\usepackage{amsmath}
\usepackage{nccmath}
\usepackage{txfonts}
\newcommand{\oao}{OAO~1657-415\xspace}
\newcommand{\nustar}{{\sl NuSTAR}\xspace}
\newcommand{\ff}[1]{\textcolor{red}{\textbf{#1 }}}

\begin{document}

  \title{A {\slshape{NuSTAR}} observation of the eclipsing binary system \oao: The revival of the cyclotron line}

   \subtitle{}

   \author{Enzo A. Saavedra\inst{1,2}, 
   Federico A. Fogantini\inst{1,2},
   Jorge A. Combi\inst{1,2},
   Federico Garc\'{\i}a\inst{1,2} \and
   Sylvain Chaty\inst{3,4}}

   \institute{Facultad de Ciencias Astron\'omicas y Geof\'{\i}sicas, Universidad Nacional de La Plata, Paseo del Bosque, B1900FWA La Plata, Argentina \and
   Instituto Argentino de Radioastronom\'ia (CCT La Plata, CONICET; CICPBA; UNLP), C.C.5, (1894) Villa Elisa, Buenos Aires, Argentina.\and 
   AIM, CEA, CNRS, Universit\'e Paris-Saclay, Universit\'e de Paris, F-91191 Gif-sur-Yvette, France \and
  Universit\'e de Paris, CNRS, Astroparticule et Cosmologie, F-75013 Paris, France
             }

   \date{Received; accepted}

% \abstract{}{}{}{}{} 
% 5 {} token are mandatory
 
  \abstract
   %context heading (optional)
   %{} leave it empty if necessary  
   {\oao is an accreting X-ray pulsar with a high mass companion that has been observed by several telescopes over the years, in different orbital phases.  Back in 1999, observations performed with {\sl Beppo-SAX} lead to the detection of a cyclotron-resonant-scattering feature, which has not been found again with any other instrument. A recent \nustar X-ray observation, performed during the brightest phase of the source, allows us to perform sensitive searches for cyclotron-resonant-scattering features in the hard X-ray spectrum of the source.}
   %aims heading (mandatory)
   {We aim to characterise the source by means of temporal and spectral X-ray analysis,  and to confidently search for the presence of cyclotron-resonant-scattering features.}
   %methods heading (mandatory)
   {The observation was divided into four time intervals in order to characterise each one. Several timing analysis tools were used to obtain the pulse of the neutron star, and the light curves folded into the time intervals. The \nustar spectrum in the energy range 3-79 keV was used, which was modelled with a power-law continuum emission model with a high-energy cutoff. }
   %results heading (mandatory)
   {We find the pulsations associated with the source in the full observation,  which are shifted due to the orbital Doppler effect. We show evidence that a cyclotron line at $35.6\pm2.5$~keV is present in the spectrum.  We use this energy to estimate the dipolar magnetic field at the pulsar surface to be $4.0\pm0.2\,\times\,10^{12}$~G.
    We further estimate a lower limit in the distance to \oao of $\simeq 1$~kpc. And we also find a possible positive correlation between the luminosity and the energy associated with the cyclotron line.}
   %conclusions heading (optional), leave it empty if necessary 
   {We conclude that the cyclotron line at $35.6\pm2.5$~keV is the same as that detected by {\sl Beppo-SAX}. Our detection has a significance of $\sim$~3.4$\sigma$.}

   \titlerunning{Re-detection of cyclotron absorption line in \oao}
   \authorrunning{Enzo A. Saavedra et al.}
   \maketitle
%
%-------------------------------------------------------------------

\section{Introduction}

\oao is a neutron-star high mass X-ray binary (NS-HMXB) formed by an accreting X-ray pulsar and an early massive stellar companion. The system was discovered with the {\sl Copernicus} satellite \citep{1978Natur.275..296P}. \cite{1980MNRAS.193P..49P} detected pulsations associated to the NS at $\sim$~38~s using {\sl HEAO A-2} observations. The companion star is an Ofpe/WNL \citep{2009A&A...505..281M} characterised by slow winds, high mass-loss rates and evidence of CNO-cycle. This spectral type is proposed as a transitional object between the main sequence OB stars and the Wolf-Rayet stars \citep{2009A&A...505..281M, 2012MNRAS.422..199M}. The binary system has an orbital period of $\sim$10.4 days \citep{1993ApJ...403L..33C} along with an eclipse duration of $\sim$1.7~days. The associated orbital period decay is $\-9.74\times10^{-8}$~s~s$^{-1}$ \citep{2012ApJ...759..124J}. Based on the properties of a scattered dust halo associated to the source, \citet{2000AAS...197.8405A} estimated a distance of about $7.1\pm1.3$~kpc, consistent with that estimated by \citet{2002ApJ...573..789C}, of $6.3\pm1.5$~kpc, based on the infrared counterpart of the source. \citet{2012MNRAS.422..199M} through near-infrared observations, reports a range of values in which the distance can be: 4.4~$\leq$~d~$\leq$~12 kpc. Recent analysis of {\sl Gaia} observations \citep{2018A&A...616A...1G} showed a relatively lower distance of  $2.2_{-0.7}^{+0.5}$~kpc, using ultra-precise angular parallaxes of the optical companion \citep{2020ApJ...896...90M}. This variability in values shows how problematic it can be to determine the distance from this source using conventional methods. \citet{2014MNRAS.442.2691P} suggests that \oao has characteristics of intermediate to normal supergiant systems and and exhibits supergiant fast X–ray transient (SFXT) phenomena  \citep{2006ESASP.604..165N}.                         

The X--ray spectrum of \oao shows high levels of absorption, which may be associated with being at a low
galactic latitude, or due to the large amount of circumstellar material. A similar scenario to the one reported in the source IGR J16320-4751 \citep{2018A&A...618A..61G} and more extremely in IGR J1618–4848 hosting a rare companion star of spectral type sgB[e] \citep{2020ApJ...894...86F}. It shows an Fe~$K\alpha$ fluorescence emission line at $\sim$6.4~keV and a corresponding Fe~$K\beta$ at $\sim$7.1~keV \citep{2000AAS...197.8405A, 2002ApJ...573..789C}. \citet{2019MNRAS.483.5687P} also showed, using Chandra observations, the presence of \ion{He}-like~\ion{Fe} \text{at} 6.7~keV, \ion{H}-like~\ion{Fe} \text{at} 6.97~keV, and \ion{Ni}~K$\alpha$ at 7.4~keV. This indicates a highly ionised surrounding medium, which is rare for this type of source \citep{2019NewAR..8601546K}. The location of the Fe ionisation region is likely to be within or close to the accretion radius \citep{2014BASI...42..147J}. Recently, \citet{2021JApA...42...72J}, through {\sl AstroSAT} observation in orbital phase $0.8245\pm0.1435$ of \oao, reports emission lines at 6.4~keV and 6.7~keV, with an equivalent width of $\sim$1~keV for both lines. A wind driven accretion wake might be present in the last orbital phase. 
The presence of a cyclotron absorption line in {\sl Beppo-SAX} observations at $\sim$36 keV was found by \citet{1999A&A...349L...9O}. This absorption line is associated with resonant scattering of photons by electrons in the Landau levels at $\sim10^{12}$~G in the polar caps of the NS. If the energy associated with the resonance are known, the magnetic field of the source can be calculated \citep{2017A&A...597A...3S}.

In this paper, we report results from a temporal and spectral X--ray analysis of the accreting X--ray pulsar with a high mass companion \oao using an observation carried out with the \nustar observatory. The observational data and reduction procedures are described in \hyperref[sec:data]{Section~\ref{sec:data}}. In \hyperref[sec:results]{Section~\ref{sec:results}} we present the results of the timing and spectral analyses. Discussion and conclusions are summarised in the \hyperref[sec:disc]{Section~\ref{sec:disc}} and~\hyperref[sec:concl]{Section~\ref{sec:concl}}, respectively.

%-------------------------------------------------------------------

\section{Observation and Data Analysis} \label{sec:data}

\begin{figure}
    \includegraphics[width=\columnwidth]{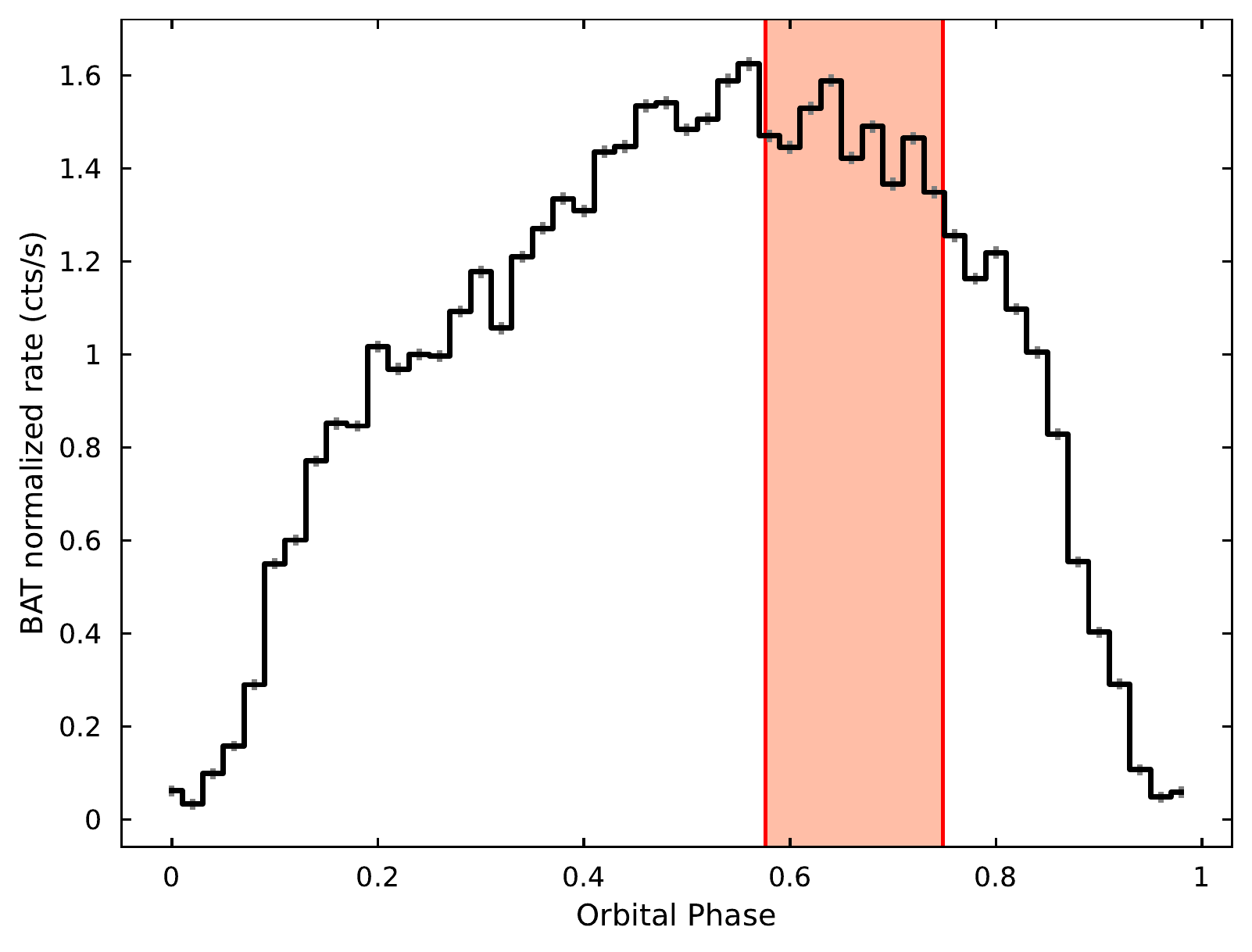}
    \caption{{\sl Swift}/BAT folded light curve of \oao using 50 phase bins, an orbital period of 10.45~days and a reference epoch of 52298.01~MJD \citep{2012ApJ...759..124J}. The vertical orange region corresponds to the \nustar observation used in this paper.
    %with its width representing the total exposure time.
    The NuSTAR observation spans $\sim$17\% of \oao's orbit, starting on phase 0.576.
    %The orbital phase covered during the observation is 0.576 to 0.748.}
    }
    \label{Fig:bat}
\end{figure}

\subsection{\nustar Data}

The \nustar telescope \citep[{\sl Nuclear Spectroscopic Telescope Array};][]{2013ApJ...770..103H} is an X--ray satellite equipped with two detectors, FPMA and FPMB, operating in the 3 to 79 keV energy range. \oao was observed in June 11, 2019 (ObsID 30401019002), with a coverage time of 154 ks and livetime of 74 ks. Data was reduced using {\tt NuSTARDAS-v. 2.0.0} analysis software from the {\tt HEASoft}~v.6.28 task package and {\tt CALDB} (V.1.0.2) calibration files. The source events were accumulated within a circular region of 125 arcseconds radius around the focal point. The chosen radius encloses $\sim90$\% of the PSF. Background events were taken from a source-free circular region with a radius of 96 arcseconds in the same CCD. The background events were detected with an average count rate of 4~c~s$^{-1}$ and a maximum count rate of 6~c~s$^{-1}$ in the 3--79~keV energy range. We extracted light curves in different energy ranges with different time groupings.

Light curves and spectra were extracted using {\tt nuproducts} task. Barycenter-corrected light curves were extracted using {\tt barycorr} task with {\tt nuCclock20100101v115} clock correction file. The coordinates for using the barycentric correction are, right ascension: 255.2037~deg and declination: -41.6557~deg. Solar system ephemerides used for barycenter correction were {\tt JPL-DE200}. Subtraction of the background light curve, and addition of the light curves of both detectors was done by means of the {\tt LCMATH} task. Corrected observations were folded to obtain a final light curve. 

XSPEC~v12.11.1 package \citep{1996ASPC..101...17A} was used to model the spectra. In order to analyse the spectral properties of the source, average and time resolved spectra were extracted for each camera. The source spectrum were rebinned to have at least 30 counts per energy bin, in the 3--79 keV energy band, in order to apply $\chi^2$ statistics. All the parameters errors are reported within the 90\% confidence region.

\subsection{{\sl Swift}/BAT Data}

The {\sl Swift}/BAT alert telescope is a transient monitor that is observing the hard X--ray sky in the 15-50 keV energy range, with a detection sensitivity of 5.3 mCrab and a time resolution of 64 s \citep{2013ApJ...770..103H}. In this work we use \oao cumulative light curve up to date, first observed on February 14, 2005. 

In \hyperref[Fig:bat]{Figure~\ref{Fig:bat}} we show the cumulative \textit{Swift}/BAT folded light curve using orbital ephemeris from \citealt{2012ApJ...759..124J}. The orange vertical stripe indicates  the corresponding orbital phase of \nustar 30401019002 observation. The width represents the total exposure time ($\phi=0.662\pm0.086$). 
It is clear that the observation was scheduled to be near the apoastron of the source as phase 0 represents mid eclipse transition.

% PERIOD = 10.44729  	    # days 
% EPOCH = 52298.01 			# MJD periastron 
% EXPOSURE = 155400 		# seconds
% TSTART = 58645.53552083 	# MJD TT
% TSTART_PHASE = 0.076272969
% DELTA_EXPO = 0.082823

%-------------------------------------------------------------------

\section{Results} \label{sec:results}

\begin{figure*}[h!] \centering
    \includegraphics[scale=0.69]{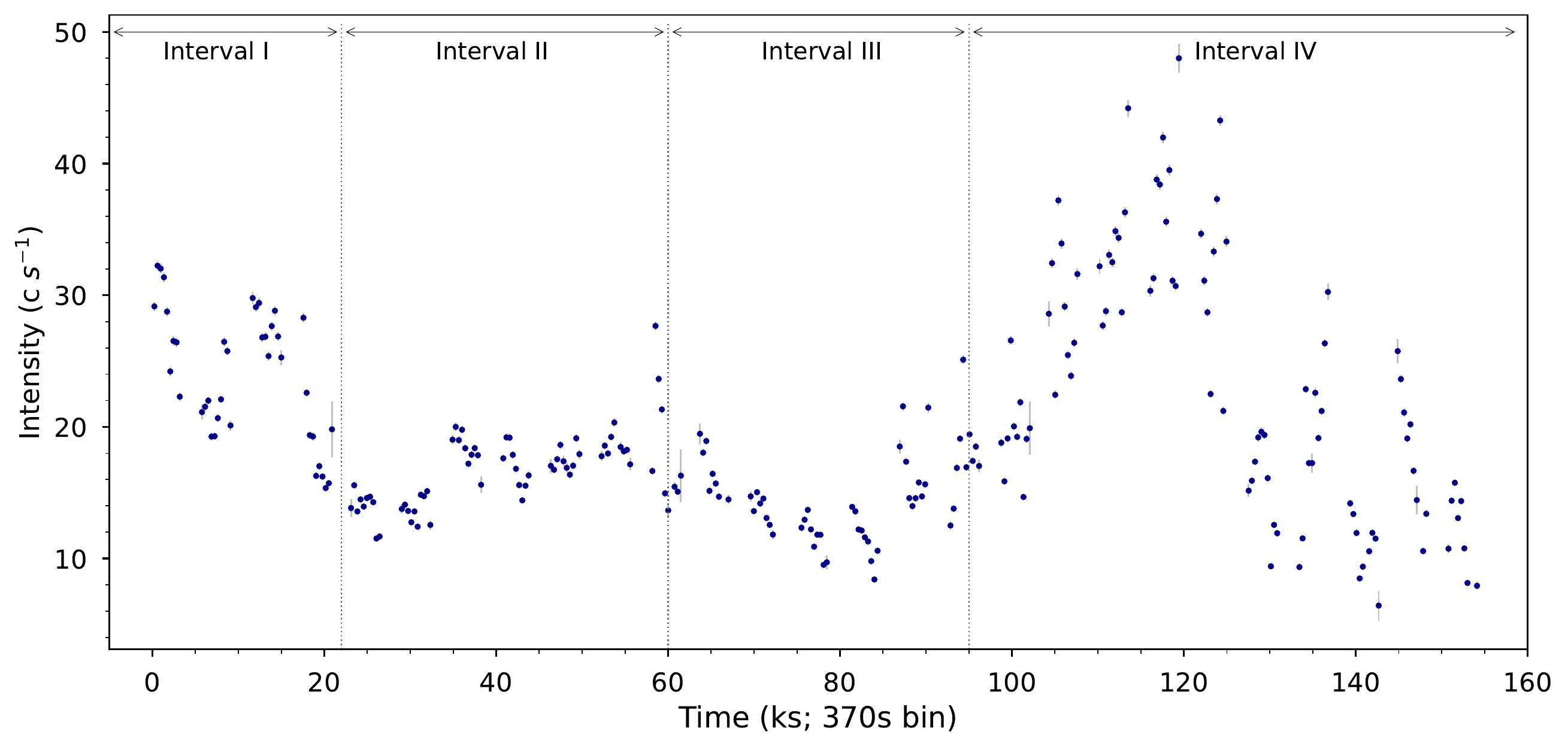}
    \caption{Background corrected light curves of \oao with a binning of 370 s, starting at 58645.5355 MJD. It was further subdivided into four intervals (colour coded) in order to make a timing and spectral analysis.}
    \label{Fig:lctotal}
\end{figure*}

\subsection{Timing analysis} \label{sec:timing}

\begin{figure}
    \includegraphics[width=\columnwidth]{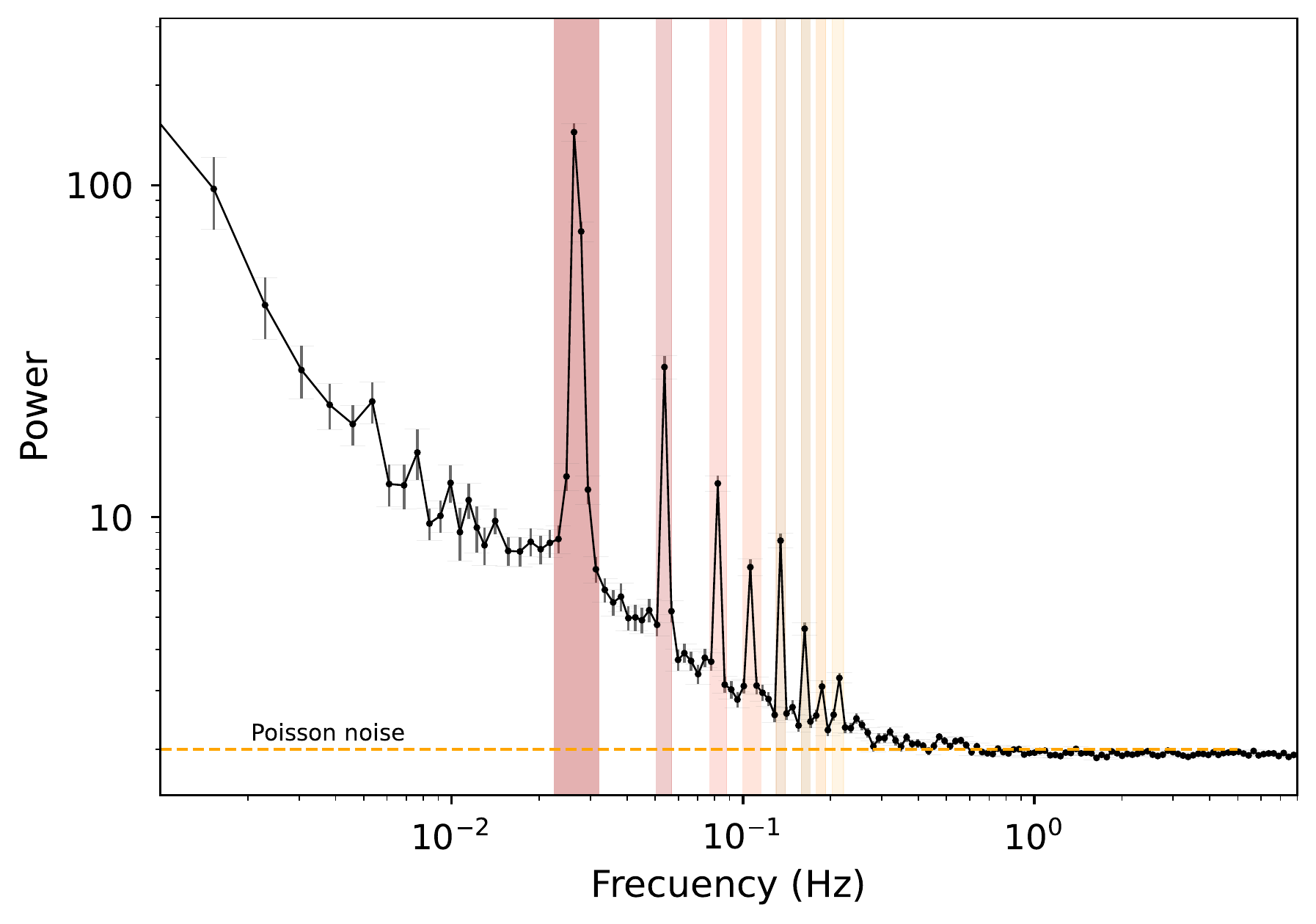}
    \caption{Power spectra associated with the complete observation. The fundamental pulse and harmonics are denoted with the vertical colour bands. The same features are seen on the remaining intervals power spectra. It was normalised using Leahy's normalisation.}
    %In all intervals the power spectra are similar.}
    \label{fig:pssI}
\end{figure}

%\begin{figure}
%    \includegraphics[width=\columnwidth]{images/loren.pdf}
%    \caption{Best period associated to interval I obtained with {\tt efsearch} task. We fit a Lorentzian distribution to the $\chi^2$ profile in order to estimate the best period uncertainty.}
%    \label{fig:loren}
%\end{figure}

In \hyperref[Fig:lctotal]{Figure~\ref{Fig:lctotal}} we show a light curve of the source with a binning of 10 times the pulsar period (370~s) in the 3--79 keV energy range. Different variability patterns can be seen, i.e., non statistical changes in count rate, ranging from a few hundred of seconds to several kiloseconds, with intensity increasing up to 5 times. We identify four time periods, shown with different colours on \hyperref[Fig:lctotal]{Figure~\ref{Fig:lctotal}}, and will be referenced from here on as Interval I,II,III and IV.

Light curves were extracted in different energy ranges associated with each interval: 3--15~keV, 15--30~keV, 30--50~keV, 50--79~keV. Using the {\tt powspec} task, we extracted power spectra associated with each interval. On \hyperref[Fig:pssI]{Figure~\ref{fig:pssI}} we show the power spectra associated with the complete observation, and on which we identify with different colours the presence of the fundamental pulse and seven harmonics. Features in the power spectra of Intervals I to IV are similar to that of the complete observation.

The {\tt efsearch} task was used to estimate the best period associated to each fundamental pulse present on each interval. 
 The period and uncertainties have been estimated by performing 1000 simulations for each light curve using the observed count statistics, as described in \cite{2013AstL...39..375B}.
In \hyperref[table:efsearch]{Table~\ref{table:efsearch}} we present the best period and uncertainties derived from the mentioned procedure for each interval. There is a decay in the pulse throughout the observation. This is possibly due to a Doppler shift in the NS spin before the eclipse, similar to what occurs in \cite{2021A&A...647A..75F}. This change in frequency is given by the following equation \citep{2005PhRvL..95z1101G}: 

\begin{ceqn}
\begin{align}
    \frac{\Delta\nu}{\nu} = \frac{x{\rm K}\pi}{\rm P_b}\left(1 - \frac{e^2}{4}\right)
\end{align}
\end{ceqn}

where $x$  the projected semi-major axis ($x=a_p\sin(i)/c$), K is semi-amplitude, $e$ is the eccentricity and ${\rm P_b}$ is the orbital period of the binary system. Taking the orbital ephemeris from \citealt{2012ApJ...759..124J}, we obtain that $\Delta$P = $1/\Delta\nu$ $\approx$~0.1~s. This equation returns the average change in frequency along the whole orbit. As the exposure time of the \nustar observation occupies $\sim$~17\% of the orbital period. On first approximation, we can weight the total average frequency change by the observed fraction. If we multiply by the factor 0.17 we get a value of $\sim$~0.017~s variation, so we have an explanation of the period shifts given by the orbital Doppler effect.

\ff{}

\begin{table}[h!]
\centering
\resizebox{.265\textwidth}{!}{%
\begin{tabular}{@{}cc@{}}
\toprule
Interval & Best periods [s]   \\ \midrule
I  &  $37.0237\pm0.0001$ \\ 
II &  $37.0277\pm0.0006$  \\ 
III & $37.0323\pm0.0004$ \\
IV &  $37.0356\pm0.0002$  \\ \bottomrule
\end{tabular}}
\caption{Best periods and uncertainties for each time interval derived from the algorithm described in \citet{2013AstL...39..375B}}
\label{table:efsearch}
\end{table}

\begin{figure}
    \includegraphics[width=0.95\columnwidth]{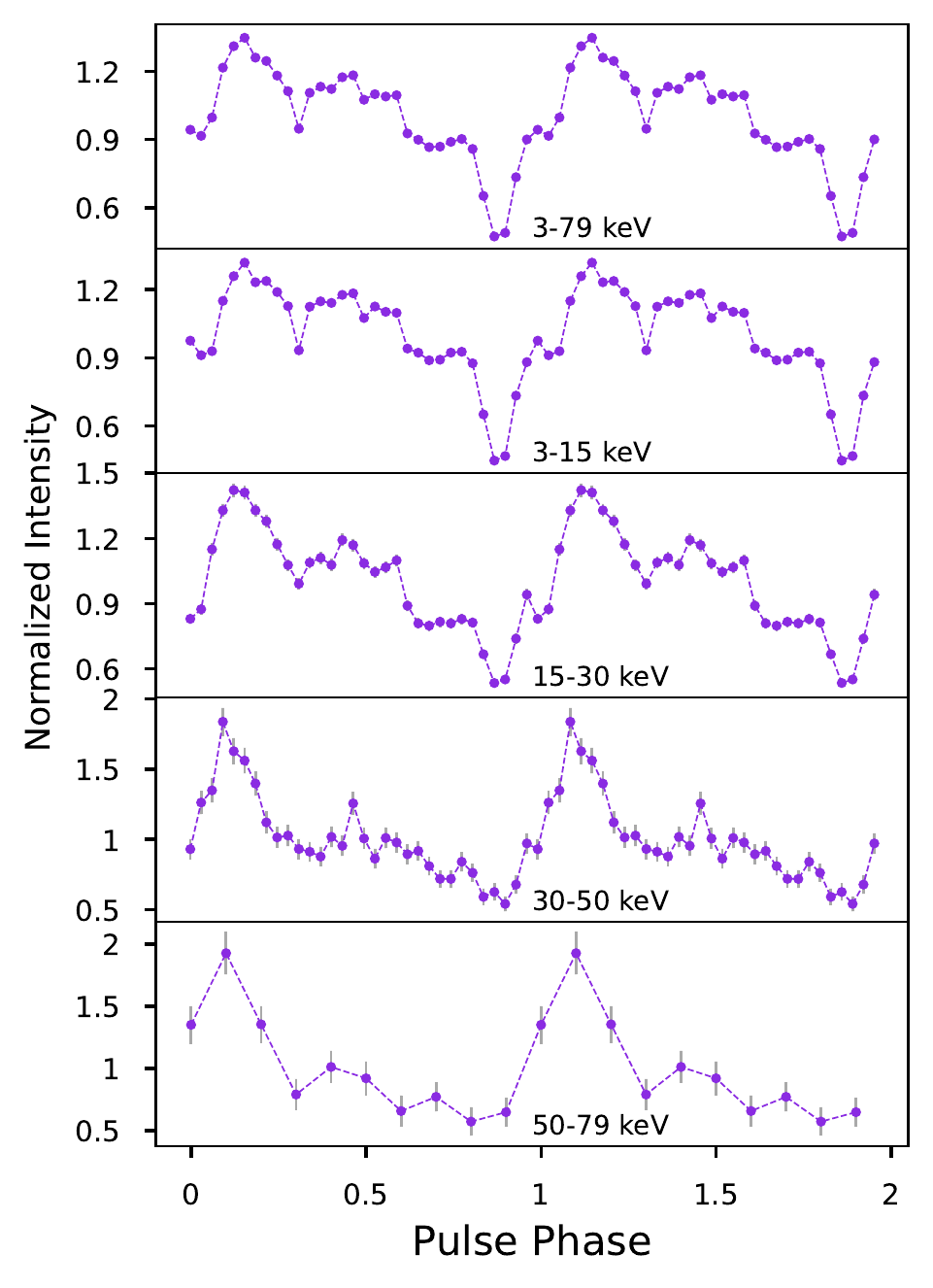}
    \caption{ Background-corrected energy-resolved pulse profiles for interval I, folded with the best period found using {\tt efsearch}}. The NS spin period modulation is detected at all the energy bands.
    \label{fig:efold}
\end{figure}

With the {\tt efold} task, we proceeded to fold light curves associated to each interval in different energy ranges. An example, associated to interval I is showed in \hyperref[fig:efold]{Figure~\ref{fig:efold}}, the other intervals are similar.  Given this analysis, pulses are detected up to the last defined interval associated with the maximum energy allowed by \nustar. It is probable that pulses are detected above 79 keV in all periods as already reported using {\sl INTEGRAL} observations \citep{2008A&A...486..293B}. 
%This shows that the pulsations are detected up to $\sim$80 keV in all intervals.       

\begin{figure}[h!]
\centering
    \includegraphics[width=\columnwidth]{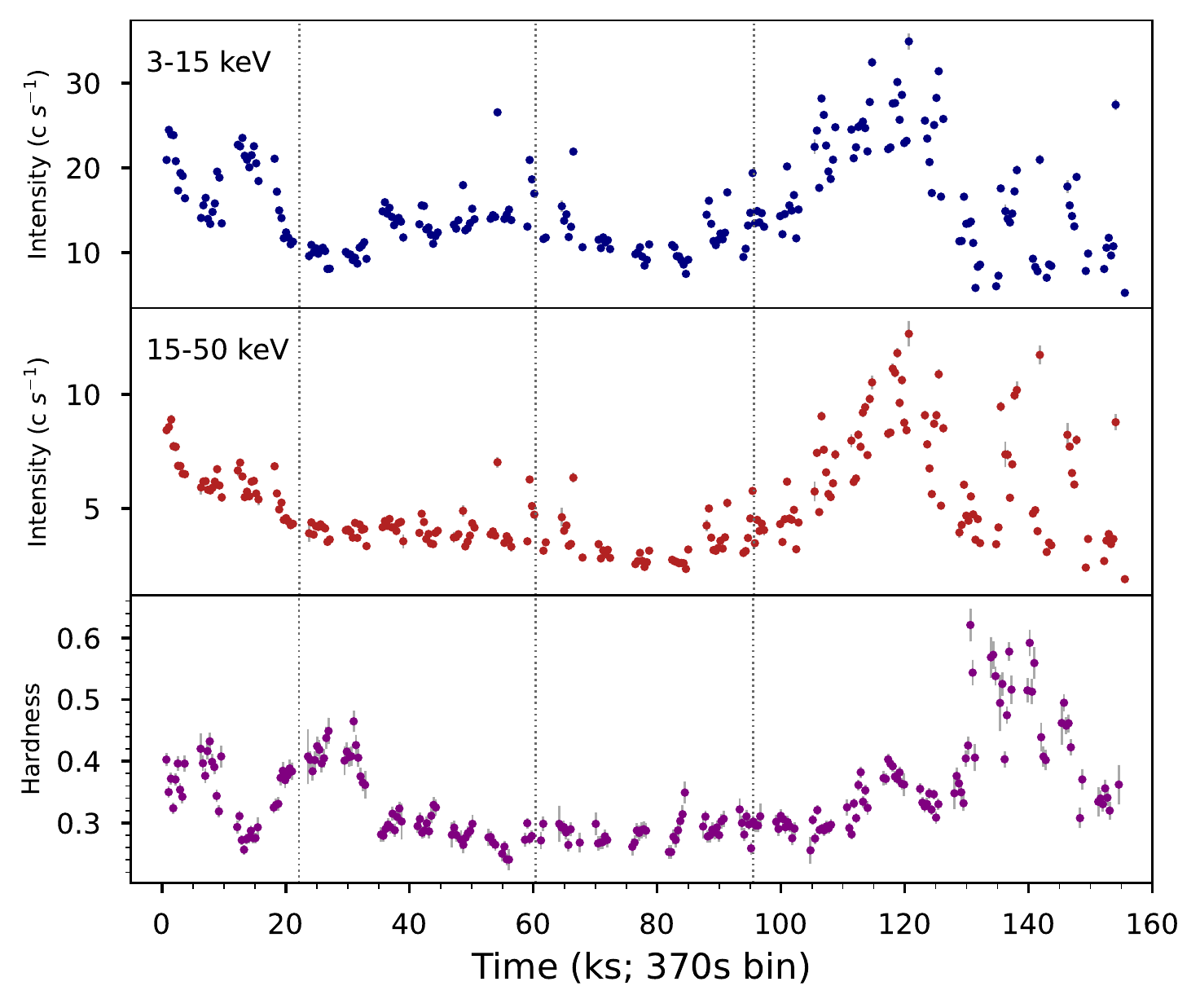}
    \caption{\nustar light curve with 370s bin size. Each plot contains 3 panels: \textit{soft} (upper panel) and \textit{hard} (middle panel) light curves, and hardness ratio (lower panel). }
    \label{Fig:hr}
\end{figure}

Light curves were also extracted for the complete observation at energy ranges: \textit{soft}: 3--15~keV and \textit{hard}: 15--50~keV. \hyperref[Fig:hr]{Figure~\ref{Fig:hr}} shows the hardness ratio obtained from these ranges. As can be seen, the brightness variations are followed by changes in the hardness ratio. 

%The light curve of each interval in the 3--79 keV energy band with a time bin of 0.1~s was used to search for pulsations around the known periodicity of 37~s with {\tt efsearch} \citep{1987A&A...180..275L}. 

\subsection{Spectral Analysis} \label{sec:spectral}

\begin{figure}[h!] 
\centering
    \includegraphics[width=\columnwidth]{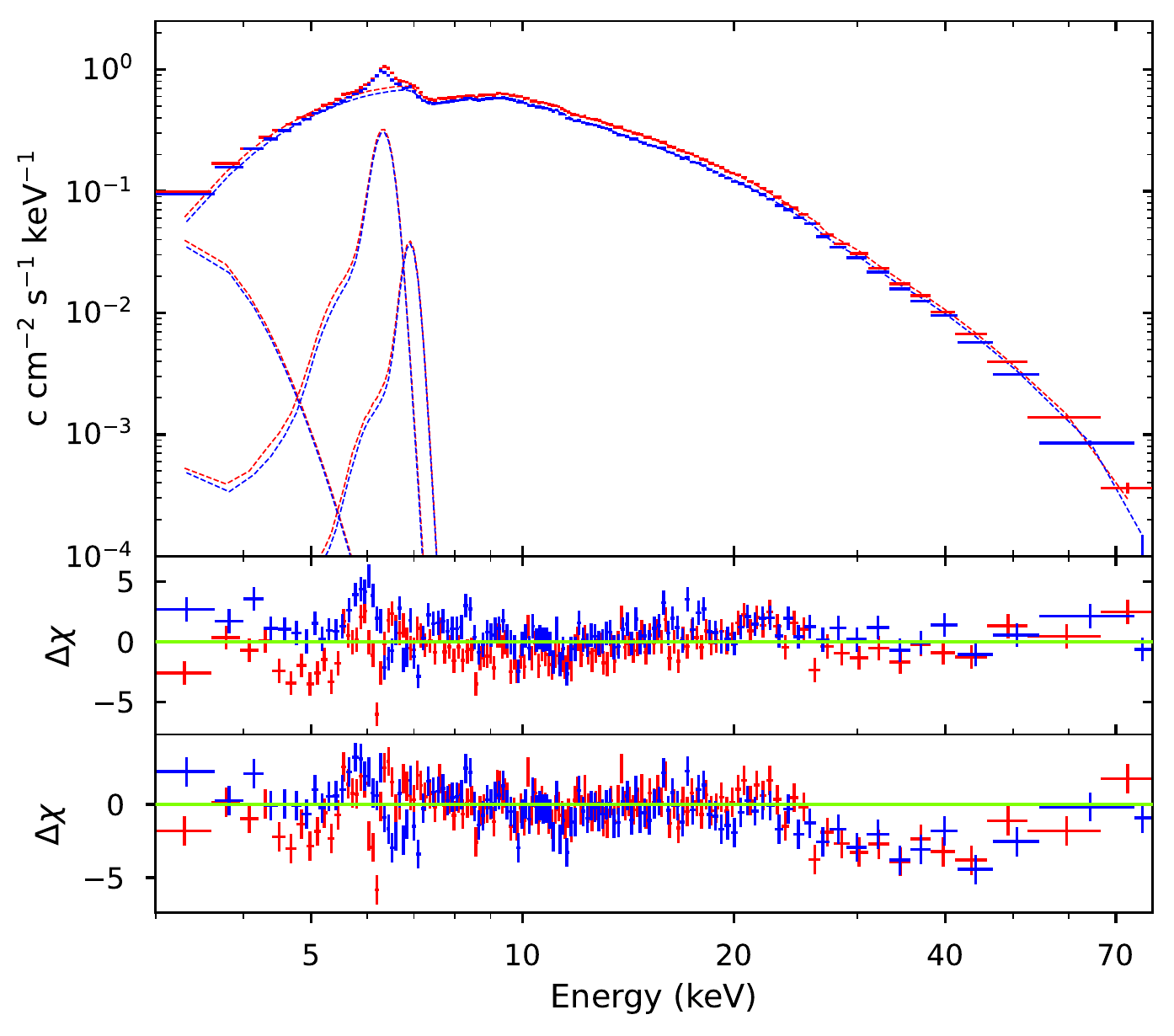}
    \caption{ {\it Top panel}: NuSTAR's FPMA and FPMB average energy spectrum of \oao. 
    The continuum can be described by a contribution of absorbed cutoff powerlaw and blackbody.
    Gaussian distributions were used to model emission lines seen at $\sim$6.4 keV and $\sim$7 keV. 
    {\it Middle panel}: residuals associated with the continuum model. 
    {\it Bottom panel}: residuals associated with the continuous model plus {\tt cyclabs} with depth equal to zero. A clear count flux deficit can be seen at around $\sim$35~keV. }
    \label{fig:rebin}
\end{figure}

%The source and background were adjusted simultaneously with the two \nustar detectors. A calibration constant was included which varies the FPMA and FPMB spectra by 2\%. The constant factor was set to 1 for FPMA data and kept free of variation for FPMB data. 

In order to characterise the spectrum of the source, we performed a spatially resolved spectral analysis, where the source and background were modelled simultaneously using the two \nustar detectors. A calibration constant was included which varies the FPMA and FPMB spectra by 2\%. 
%The constant factor was set to 1 for FPMA data and kept free of variation for FPMB data. 

%, different models were tested using {\sc XSPEC}. 

The source broad-band spectrum in NS-HMXBs has a quite complicated shape and can be appropriately fitted by a composite model with two continuum components: a  blackbody emission, with the {\it kT} at low energies and a power law with an exponential cut-off at high energies. 
In particular, for OAO~1657$-$415, the spectrum was fitted using different models for the high energy component: {\tt highecut}, {\tt bknpow}, {\tt cutoffpl}, {\tt comptt} \citep{1994ApJ...434..570T}. 
After several tests, we found that the best fit is consistent with a {\tt cutoffpl}. 
For the lower energetic component we tested the {\tt bbody} and {\tt diskbb} models \citep{1984PASJ...36..741M}. This ensures that the emission comes from the neutron star or the accretion disk. The best fit was obtained with a {\tt bbody} at $\sim$0.2 keV. 
The interstellar absorption was modelled using the Tuebingen-Boulder interstellar absorption model ({\tt tbabs}), with solar abundances set according to  \citet{2000ApJ...542..914W}, and the effective cross sections given by \citet{1996ApJ...465..487V}.
Two emission lines are present at $\sim$6.4 keV and $\sim$7 keV \citep{2000AAS...197.8405A, 2002ApJ...573..789C}. $K{\beta}$ emission line width was left frozen to the best fit value. 
The final model results in a $\chi^{2}/{\rm dof}$: 2434.83/2249.

\begin{table*}
  \begin{center}
    \begin{tabular}{lllll}
      \toprule 
       \textbf{Component}        & \textbf{Parameters}        &  \textbf{Model I} & \textbf{Model II}  & \textbf{Model III}\\
      \midrule 
      \textit{CONST}  & $C_{FPMA/FPMB}$              & $1.020\pm0.003$ & $1.020\pm0.003$  & $1.020\pm0.003$ \\
        &  &  &  &  \\
     \textit{TBABS}  & $N_{H}~(\rm 10^{22}~cm^{-2})$        & $26.4\pm0.9$   & $26.7\pm0.9$ & $26.7\pm1$\\
        &  &  &  &  \\
     \textit{CYCLABS} & $E_{cyc}~(\rm keV)$             &   $-$    &   $35.3_{-5.8}^{+3.0}$    &   $35.6\pm2.5$  \\
       & $\sigma_{cyc}~(\rm keV)$              &   $-$             &  $24.9\pm12.5$          &   $10^\dagger$  \\
      & $Depth_{cyc}$               &  $-$                         &   $0.07\pm0.03$           &   $0.07\pm0.03$ \\
       &  &  &  &  \\
    \textit{CUTOFFPL} & $\Gamma$                    &  $0.54\pm0.03$   & $0.63\pm0.06$ & $0.56\pm0.04$\\
     & $E_{cut}~(\rm keV)$          & $16.4\pm0.4$  & $19.9_{-2.4}^{+3.8}$ & $17.4\pm0.7$ \\
     & ${\it Norm_{cpl}}$  & $0.014\pm0.001$ & $0.016\pm0.001$ & $0.015\pm0.001$\\
       &  &  &  &  \\
    \textit{BBODY} & $kT_{bb}~(\rm keV)$      & $0.22\pm0.01$ & $0.22\pm0.01$ & $0.22\pm0.03$\\
     & ${\it Norm_{bb}}$  & $0.5\pm0.3$  & $0.7\pm0.4$  & $0.6\pm0.4$\\
      &  & &  &   \\
    \textit{GAUSS} & $E_{K\alpha}~(\rm keV)$                    & $6.33\pm0.007$  & $6.330\pm0.01$  & $6.33\pm0.01$ \\
     & $\sigma_{K\alpha}~(\rm keV)$              & $0.10\pm0.01$   & $0.10\pm0.04$  & $0.10\pm0.04$ \\
     &  Norm~($10^{-4}$)   & $9.0\pm0.3$   & $9.0_{-0.2}^{+0.6}$  &  $9.0_{-0.2}^{+0.6}$ \\
      & $E_{K\beta}~(\rm keV)$                     & $6.91\pm0.03$ & $6.9\pm0.03$ & $6.91\pm0.03$ \\
     &  $\sigma_{K\beta}~(\rm 10^{-3}~keV)$               & $1.47^\dagger$   & $7.1^\dagger$   & $1.62^\dagger$\\
     &  Norm ($10^{-4}$)   & $7.0\pm0.1$  & $7.0\pm0.3$  &  $8.0\pm0.3$\\
     \midrule
    &  $\chi^{2}/{\rm dof}$                        &   $2434.83/2249$  &  $2398.51/2247$ &  $2393.87/2246$ \\ 
      \bottomrule 
    \end{tabular}
  \end{center}
  \caption{Final parameters of the average energy spectrum associated with the described models: Model I:  {\tt bbody}+{\tt cutoffpl}+{\tt gauss}; Model II: same as Model I plus {\tt cyclabs}; Model III: same as Model II but with $\sigma_{cyc}$ fixed at 10 keV. \\
  $^\dagger$~Indicates that the parameter was frozen before error calculations. }
  \label{table:speca}
\end{table*}

\begin{table*}
  \begin{center}
    \begin{tabular}{llllll}
      \toprule 
     \textbf{Component}  & \textbf{Parameters}        &  \textbf{Interval I} & \textbf{Interval II}  & \textbf{Interval III} & \textbf{Interval IV}\\
      \midrule 
    \textit{CONST}  & $C_{FPMA/FPMB}$              & $1.020\pm0.007$ & $1.020\pm0.007$  & $1.020\pm0.008$ &  $1.020\pm0.004$ \\
     & &  &  & &  \\
      \textit{TBABS}  & $N_{H}~(\rm 10^{22}~cm^{-2})$          & $28.9\pm2.8$ & $26.9\pm1.8$  & $ 25.6\pm2.2$ &  $26.8\pm1.5$ \\
     & &  &  & &  \\
     \textit{CYCLABS} & $E_{cyc}~(\rm keV)$            & $38.2\pm4.5$ &   $36.7\pm3.5$   & $28_{-27}^{+17}$ &    $31.3\pm2.6$     \\
      & $Depth_{cyc}$              &$0.14\pm0.07$  &  $0.12\pm0.05$     &  $0.028\pm0.04$ &    $0.062\pm0.02$     \\
     &  &  &  & & \\
      \textit{CUTOFFPL} & $\Gamma$       & $0.62\pm0.1$     &  $0.73\pm0.08$ & $0.83\pm0.09$ & $0.43\pm0.06$          \\
     & $E_{cut}~(\rm keV)$ & $18.5_{-1.6}^{+2.1}$     & $19.4_{-1.5}^{+1.7}$  & $19.2_{-1.5}^{+2.4}$ &  $16.10\pm0.72$     \\
     & ${\it Norm_{cpl}}$     & $0.020\pm0.005$  & $0.020\pm0.003$ &  $0.020\pm0.005$ &  $0.010\pm0.002$  \\
     &  &  & &  &   \\
     \textit{BBODY} & $kT_{bb}~(\rm keV)$ & $0.24\pm0.02$     & $0.20\pm0.02$  & $0.20\pm0.02$ & $0.240\pm0.002$ \\
     & ${\it Norm_{bb}}$     & $0.5_{-0.3}^{+0.8}$      & $2.3_{-1.6}^{+8.1}$  &  $1.5_{-1.1}^{+8.6}$ &  $0.4_{-0.2}^{+0.5}$ \\
      &  &   & &\\
      \textit{GAUSS} & $E_{K\alpha}~(\rm keV)$        & $6.33\pm0.01$           & $6.35\pm0.01$ & $6.3\pm0.01$ & $6.33\pm0.009$ \\
    &  $\sigma_{K\alpha}~(\rm keV)$  & $0.08\pm0.04$           & $0.10\pm0.03$  &  $0.09\pm0.04$ & $0.10\pm0.02$ \\
     & Norm ($10^{-4}$)   & $10.0\pm0.1$     & $8.0\pm0.3$ & $7.0_{-0.6}^{+0.4}$ & $1.0\pm0.6$ \\
                      & $E_{K\beta}(\rm keV)$         & $7.04\pm0.2$           & $6.91\pm0.1$  & $6.99\pm0.4$ & $6.8\pm0.07$ \\
    &  $\sigma_{K\beta}~(\rm 10^{-3}~keV)$   & $140^\dagger$                   & $6.1^\dagger$    & $0.2^\dagger$  &  $3.2^\dagger$ \\
     & Norm ($10^{-4}$)  & $1.0\pm0.3$     &$6.0\pm0.3$  & $5.0\pm0.3$ & $4.0\pm0.3$\\
    \midrule
    &  $\chi^{2}/{\rm dof}$          &  $1586.31/1459$                &   $1491.58/1491$  & $1376.86/1344$   & $1838.39/1861$ \\ 
      \bottomrule 
    \end{tabular}
  \end{center}
  \caption{Parameters associated with the models in different time intervals. $\sigma_{cyc}$ was fixed at 10~keV in the three associated intervals. \\
  $^\dagger$~Indicates that the parameter was frozen during setting. }
  \label{table:specregion}
\end{table*}

\raggedbottom

\begin{figure*}[h!]
     \centering
     \begin{subfigure}[b]{0.48\textwidth}
         \centering
         \includegraphics[width=\textwidth]{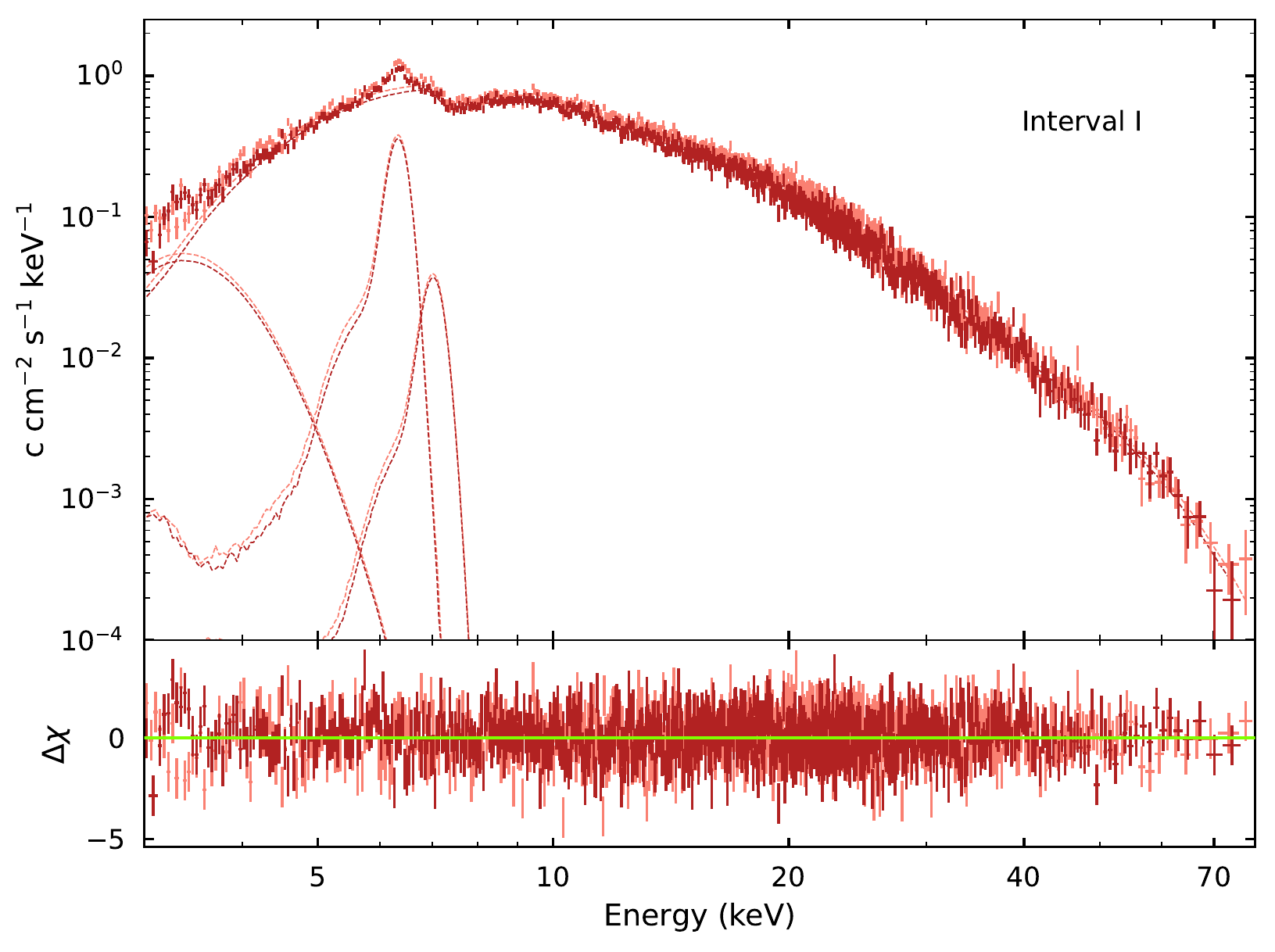}
         \caption{}
         \label{fig:psI}
     \end{subfigure}
     \hfill
     \begin{subfigure}[b]{0.48\textwidth}
         \centering
         \includegraphics[width=\textwidth]{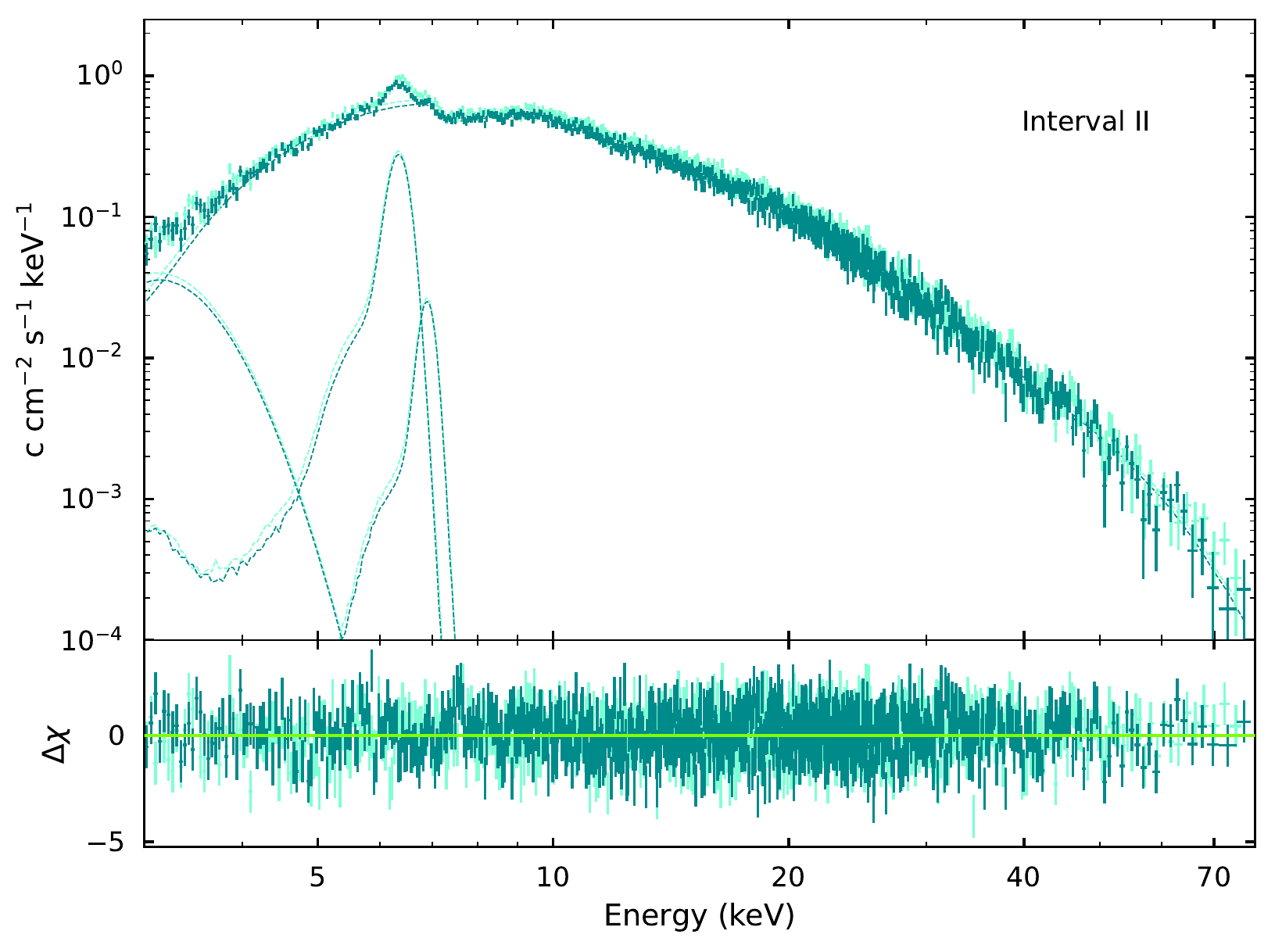}
         \caption{}
         \label{fig:psII}
     \end{subfigure}
     \hfill
     \begin{subfigure}[b]{0.48\textwidth}
         \centering
         \includegraphics[width=\textwidth]{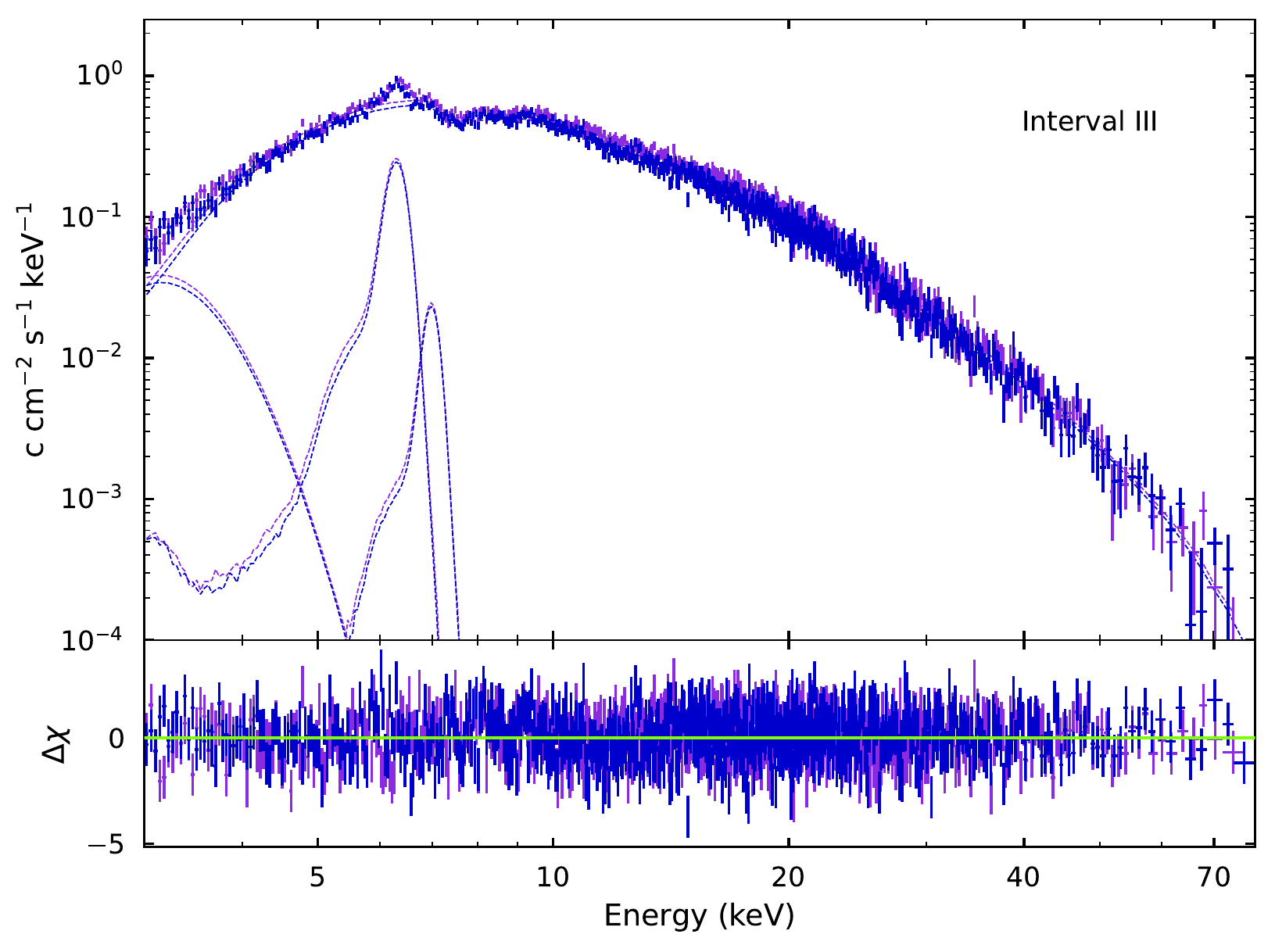}
         \caption{}
         \label{fig:psIII}
     \end{subfigure}
          \hfill
     \begin{subfigure}[b]{0.48\textwidth}
         \centering
         \includegraphics[width=\textwidth]{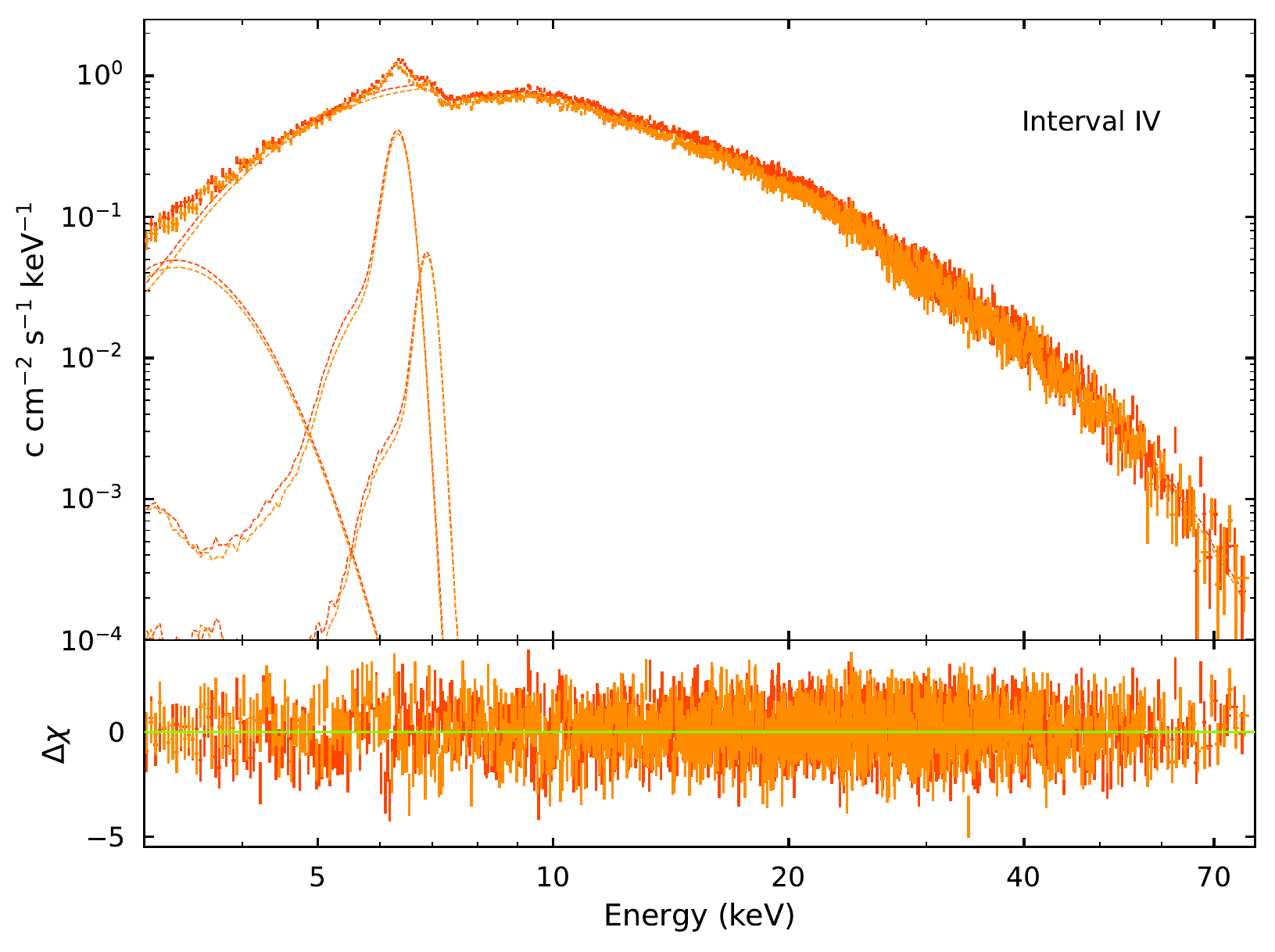}
         \caption{}
         \label{fig:psIV}
     \end{subfigure}
        \caption{Time resolved spectra for each defined interval fitted with Model {\tt bbody}+{\tt cutoffpl}+{\tt gauss} plus {\tt cyclabs}. The bottom panel of each subplot indicates the model residuals.}
        \label{fig:specregion}
\end{figure*}

%\begin{figure*}[h!]
%\centering
%    \includegraphics[width=2.\columnwidth]{images/cyclabs.png}
%    \caption{One- and two-dimensional projections of the probability distributions derived from the MCMC analysis for the parameters of the continuous model. This corresponds to the average energy spectrum associated with Model {\tt bbody}+{\tt cutoffpl}+{\tt gauss} plus {\tt cyclabs} with $\sigma_{\rm cyc}$ fixed at 10 keV. The contours in the two-dimensional projections for the parameters correspond to the 1-, 2- and 3-$\sigma$ confidence interval.}
%    \label{Fig:MCMC}
%\end{figure*} 

We included a Lorentzian cyclotron line component {\tt cyclabs} to the model \citep{1990Natur.346..250M, 1990ApJ...365L..59M}. A count flux deficit can be seen around $\sim$35 keV (see \hyperref[fig:rebin]{Figure~\ref{fig:rebin}}) if we fix depth to zero. We obtain a very large value of cyclotron line width, $\sigma_{cyc}\sim$25keV, which is statistically equal to the value obtained by \citet{1999A&A...349L...9O} using {\sl Beppo-SAX} observations. 
For subsequent fits we left the line width fixed at 10~keV. This is consistent with the correlation between the width and the energy centroid of the cyclotron lines for different X--ray binaries \citep{2020A&A...642A.196S}. 
The final fit gives  $\chi^{2}/{\rm dof}$: 2398.51/2247, which is significantly better than the model without the cyclotron absorption component.

Final model parameters and uncertainties are listed in  \hyperref[table:speca]{Table~\ref{table:speca}}. The {\tt cutoffpl} normalisation reported is in units of photons~keV$^{-1}$~cm$^{-2}$~s$^{-1}$ at 1~keV. The {\tt blackbody} normalisation reported is in units of ${L_{39}}/{D_{10}^2}$ where $L_{39}$ is the source luminosity in units of $10^{39}$~erg~s$^{-1}$ and $D_{10}$ is the distance to the source in  units of 10~kpc.  The {\tt Gaussian} normalisations reported are in units of total photons~cm$^{-2}$~s$^{-1}$ in the line. 
The equivalent widths of the K$\alpha$ and K$\beta$ emission lines are $0.257\pm0.01$~keV and $0.0248\pm0.009$~keV respectively. \citet{2021JApA...42...72J} reports an equivalent width of 1 keV. This value does not differ from the one found here, as it is poorly constrained, and depends on the observation and segments considered.
%These values differ significantly from those reported by \citet{2021JApA...42...72J}, which are $\sim$1~keV.
The unabsorbed flux was computed using the {\tt cflux} convolution model, yielding $1.17\times10^{-9}$ erg~cm$^{-2}$~s$^{-1}$ in the 3--79 keV energy range , for the total continuum spectrum and the emission lines inclusively.                                                                                      
%As the cyclotron absorption line is in the data but is weak, pulse-phase resolved spectroscopy could not be performed. However, it was possible to analyse its presence in the intervals defined above. 

To analyse the spectral characteristics of each time interval, the spectra associated with these were extracted. The four periods were described with a {\tt cutoffpl}+{\tt bbody} model for the continuum and two Gaussian distributions to account for the Fe emission lines. 
In addition, a Lorentzian cyclotron line component was included in three of the four intervals. Interval III could not be correctly fitted with the addition of the cyclotron component, so the centroid energy line had to be fixed at best fit, which was $\sim$34 keV.
%which could be due to insufficient statistics. 
The final model parameters for each interval are listed in  \hyperref[table:specregion]{Table~\ref{table:specregion}}, and their associated spectra shown in \hyperref[fig:specregion]{Figure~\ref{fig:specregion}}. 
The centroid energy $E_{cyc}$ can vary between 31.3 keV and 38.2 keV.

\begin{table}[h!]
\centering
\begin{tabular}{@{}lccccc@{}}
\toprule
Interval & I & II & III & IV \\ 
$\sigma_{cyc} = 1-r$ (\%) & $99.96$ & $97.67$ & $30.18$ & $16.65$ \\ \bottomrule
\end{tabular}
\caption{Cyclotron absorption line significance, $\sigma_{cyc} = 1-r$, for each time interval. See text for details.}
\label{table:ratio}
\end{table}

\raggedbottom

To test the significance of the cyclotron absorption line we use the {\tt F-test} routine in {\tt XSPEC} by comparing the model {\tt bbody}+{\tt cutoffpl}+{\tt gauss} with the same model with the additional cyclotron absorption line. This results in an F-statistic of 17.01 and a probability of $4.6\times10^{-8}$. However, this analysis is not entirely correct \citep{1999A&A...349L...9O, 2002ApJ...571..545P}. It was decided to use another strategy: simulations were performed with the {\tt fake-it} task of {\tt xspec}. The line energy and depth was adjusted,  leaving fixed width at 10~keV. With this, we drew $5\times10^4$ sampled spectra with identical model and continuum parameters, and then counted how many of the sampled spectra had a depth greater than depth of the adjusted line of the real spectra. The ratio, $r$, between the latter number and the total sampled spectra gives an estimate of the probability of obtaining a depth higher than the actual spectra just by chance. The lower this ratio, the more confident we can assume the real presence of a cyclotron absorption line. The results of this evaluation for the total observation and each time interval chosen are presented in \hyperref[table:ratio]{Table~\ref{table:ratio}}.  We note that the presence of the line decreases with the passage of observation time. This is to be expected as the NS is eclipsed. We consider the presence of the cyclotron line to be significant in interval I and II, with a significance of 99.96\% and 97.67\%, respectively. For intervals III and IV we consider that the presence is not significant.  It is unusual that the width of the absorption line is very large since the range of widths in the literature varies from 0 to 15 keV. In any case larger widths are reported, except for \citet{1999A&A...349L...9O}. We consider that the absorption line is weak, and therefore the width is affected by this. This can be reflected in the large errors associated with the width and its decreasing depth over time.

The obtained differences for the model parameters between intervals are typical for many NS-HMXBs  \citep[i.e.,][]{10.1093/mnras/stab1862}. These differences give us information about the relationship between the geometry of the pulsar beam and the magnetic field \citep{2017JApA...38...50M}.

%Luminosity}
%average:6.81\times10^{35}$
%regI: $8.759\times10^{35}$
%regII: $6.01\times10^{35}$
%regIII: $5.03\times10^{35}$
%regIV: $8.23\times10^{35}$
%-------------------------------------------------------------------

\section{Discussion}
\label{sec:disc}

\begin{figure}[h!]
\centering
    \includegraphics[width=\columnwidth]{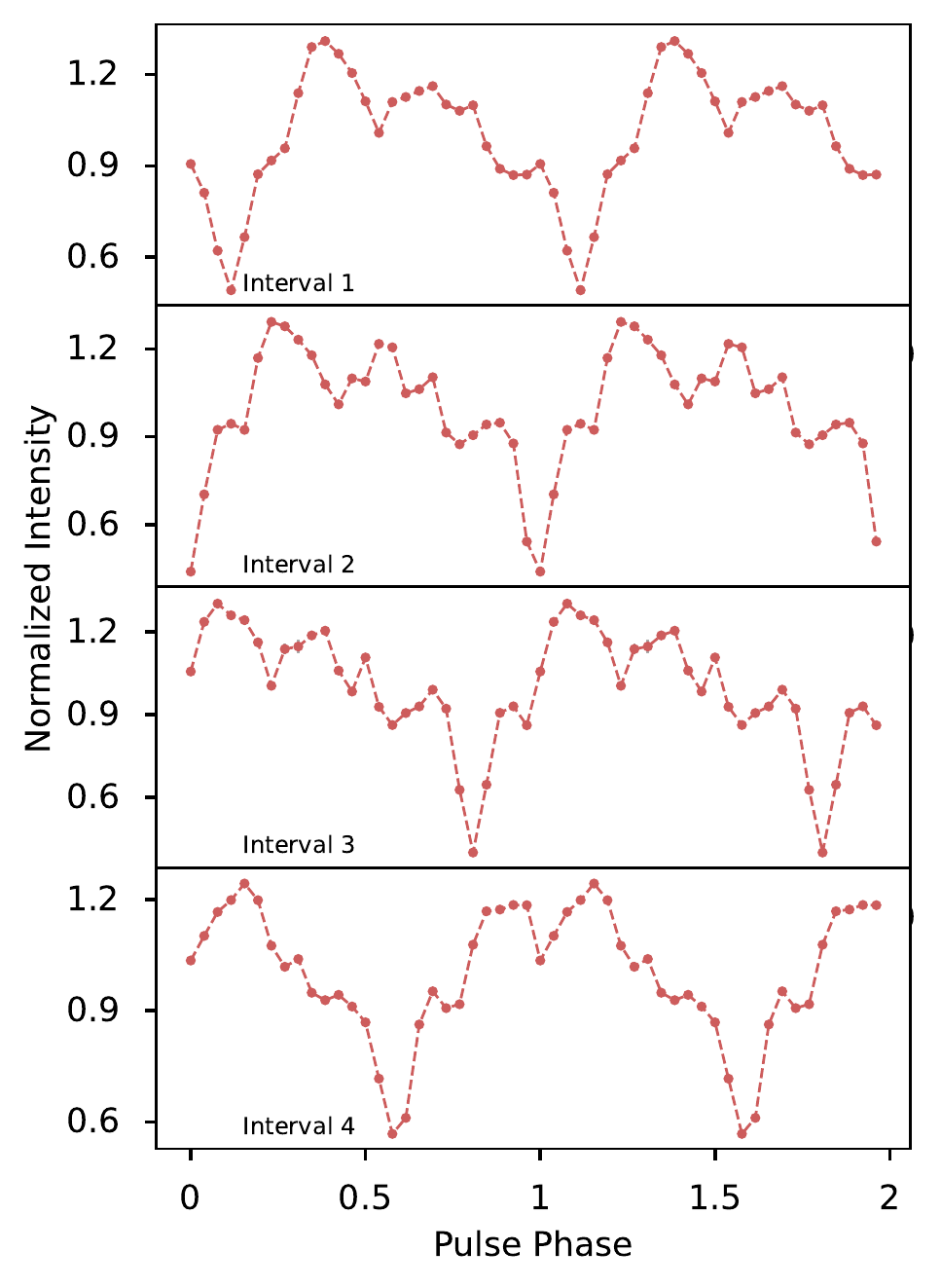}
    \caption{ Background corrected pulse profiles associated with each interval folded with the best periods found (see \hyperref[table:efsearch]{Table~\ref{table:efsearch}}).}
    \label{fig:efoldtotal}
\end{figure}

\hyperref[fig:efoldtotal]{Figure~\ref{fig:efoldtotal}} shows the folded light curves associated with each time interval, in the 3--79~keV energy range. It was obtained using light curves folded at the intervals shown in \hyperref[table:efsearch]{Table~\ref{table:efsearch}}. The epoch for all pulsations was defined at  58645.5 MJD, in order to be able to compare them. As can be seen, all the pulsations are out of phase. The observed phase shift is due to the different pulse periods adopted for the epoch folding in each interval. Different time folding using the same epoch but different pulses introduces a bogus pulse phase shift:
\begin{ceqn}
\begin{align}
\Delta \Phi = \frac{\Delta\text{P}~\Delta\text{t}}{\text{P}}
\end{align}
\end{ceqn}

where $\Delta \text{t}$ is the time distance between the the midpoints of the two time intervals and $\Delta \text{P}$ is the two periods difference. The phase shift of the pulses observed in \hyperref[fig:efoldtotal]{Figure~\ref{fig:efoldtotal}} is compatible with the formula above. The pulsations of intervals I, II and III are similar to each other. The pulse in interval IV is the most unusual, it also has less intensity than other pulsations. Moreover, the spectral index drops noticeably by half with respect to interval III. This is accompanied by a change in the hardness ratio diagram, as shown in   \hyperref[Fig:hr]{Figure~\ref{Fig:hr}}. The light curve of this interval presents a flare feature. %It is likely that these pulse offsets are due to the adjustment of the pulsar due to accretion, which is highly variable as it is driven by the wind. In every interval the temperature associated with the blackbody component remains constant, within errors, at $\sim$0.22~keV. 

%At the same time, it can be seen from  \hyperref[table:efsearch]{Table~\ref{table:efsearch}} that the total pulse variation in the NuSTAR observation gives and increase of 0.011~s. This pulse shift on the spin period can be explained by the Doppler effect given by the orbital motion of the NS, which is exiting the eclipse \citep[i.e.,][]{2021JApA...42...58S}.
%This is because the observed spin period is modulated by the Doppler shift derived from its orbital motion .

\subsection{Cyclotron resonant scattering features}

Cyclotron Resonant Scattering Features (CRSFs) or cyclotron absorption lines are produced near the magnetic poles of an accreting neutron star. 
Electrons move perpendicularly to the magnetic field lines, assuming discrete energy levels referred as Landau levels. 
Resonance features are generated when photons are scattered by electrons at resonance energy, which is commonly seen in the X-ray spectrum as a deficit in the continuum flux at energies between 10 and 80~keV \citep{2017symm.conf..153J}. 
The energy of the fundamental level corresponds to the energy gap between two adjacent Landau levels. 
The detection of this line can be used to estimate the  dipolar surface magnetic field of pulsar \citep{2018RAA....18..142C}:

%One can describe through Landau quantization levels the motion of electrons moving cyclically through a strong magnetic field (near the pulsar surface). As a consequence, the plasma becomes optically thick, generating a scattering of photons, which we will observe as an absorption in the X--ray spectrum. This absorption is what we call Cyclotron Resonant Scattering Features (CRSFs) or cyclotron absorption line. The observation of this line helps us to calculate the dipolar surface magnetic field of pulsar \citep{2018RAA....18..142C}:

\begin{ceqn}
\begin{align}
\left(\frac{B_\mathrm{cyc}}{10^{12}~\mathrm{G}}\right) = \frac{1+z_g}{n}~\left(\frac{E_\mathrm{cyc}}{\mathrm{11.6~keV}}\right)
\end{align}
\end{ceqn}

%\begin{ceqn}
%\begin{align*}
%    {\rm E_{CRSF}}=\frac{\rm E_{Landau}}{\rm 1+z_g}\approx~\frac{11.6}{\rm 1+z_g}\times~{\rm B_{12}~keV}
%\end{align*}
%\end{ceqn}

\noindent Using $E_\mathrm{cyc} = 35.6\pm2.5$~keV as the fundamental energy ($n=1$), and assuming a canonical gravitational redshift for the NS, of $z_g=0.306$ \citep{2019RAA....19..146C}, we estimate a dipolar surface magnetic field of $B_\mathrm{cyc} \approx 4.0\pm0.2\,\times\,10^{12}$~G, consistent with other NS-HMXB sources \citep{2017symm.conf..153J, 2019A&A...622A..61S}.

\subsection{Distance to \oao}

From the derived dipolar magnetic field intensity, we can estimate the distance to \oao through the equation \citep{1997ApJ...482L.163C}:

% \begin{multline*}
%     \mathrm{d~[kpc]}=\frac{\mathrm{B~P}^{-\frac{7}{6}}}{4.8\times10^{10}~G}~\left(\frac{\mathrm{F_x}}{10^{-9}~\mathrm{erg~cm^{-2}~s^{-1}}} \right)^{-\frac{1}{2}} \\
%   \left(\mathrm{\frac{M}{1.4M\odot}}\right)^{-\frac{1}{3}}\left(\mathrm{\frac{R}{10^6~cm}}\right)^{5/2}
% \end{multline*}
    
\begin{ceqn}
\begin{align*}
    \frac{d}{\mathrm{kpc}}=\frac{B P^{-\frac{7}{6}}}{4.8\times10^{10}~\mathrm{G}}~\left(\frac{F_x}{10^{-9}~\mathrm{erg~cm^{-2}~s^{-1}}} \right)^{-\frac{1}{2}}
  \left(\frac{M}{1.4~\mathrm{M}_{\odot}}\right)^{-\frac{1}{3}}
\end{align*}
\end{ceqn}

\noindent where F$_{\rm x}$ is the minimum bolometric X-ray flux at which the X-ray pulsations are still detectable. In our case, we get $1.17\pm0.01\,\times\,10^{-9}$~erg~cm$^{-2}$~s$^{-1}$, in the 3--79~keV energy range. 
Assuming a NS mass of $1.8\pm0.3$~M$_{\odot}$ \citep{2015A&A...577A.130F}, a standard radius of 10~km and a spin period $P$ of $37.03\pm0.01$~s, we obtain an approximate distance to \oao of $1.05\pm0.08$~kpc. 
This method is not as accurate as others, given that the error in F$_{\rm x}$ can be significant, leading to an error on the distance much larger than indicated  (see \hyperref[sec:corbet]{Subsection~\ref{sec:corbet}} for possible explanation of flux variation).  Also, F$_{\rm x}$ may not be the minimum bolometric X-ray flux. \citet{2008A&A...486..293B} analysed energy-resolved pulse profiles using {\sl INTEGRAL} observations, in which they note that the pulses are detected in the 120--160~keV energy range. We will take F$_{\rm x}$ obtained in this work as an approximate upper limit. Therefore, we consider this value to be a rough lower limit compared to distances already established with other methods.
The distance obtained with {\sl Gaia} based on ultra--precise angular parallaxes of the optical companion, which is $2.2_{-0.7}^{+0.5}$~kpc, is consistent with the value obtained by this method \citep{2020ApJ...896...90M}. 
\citet{2009A&A...505..281M} report a distance using a luminosity range between ${1.5\times10^{36}}{\rm erg\,s}^{-1}$ and ${10^{37}}{\rm erg\,s}^{-1}$. The lower limit is compatible with the distance obtained with {\sl Gaia}, and therefore with  the rough lower limit reported in this work.

\subsection{Possible weak dependence between luminosity and the CRSF}

In systems where a cyclotron absorption line appears there may be a positive or negative correlation between cyclotron line energy $E_{cyc}$ and X--ray luminosity \citep{2019NewAR..8601546K}. 
In order to check the statistical significance of such relationship on the \nustar data, we used the average unabsorbed luminosity and the unabsorbed luminosities associated with each interval, assuming a distance to the source of $2.2_{-0.7}^{+0.5}$~kpc \citep{2020ApJ...896...90M}. 
The average 3--79 keV unabsorbed luminosity is $\sim{6.81\times10^{35}}{\rm erg\,s}^{-1}$. The unabsorbed luminosities associated with each interval are listed in \hyperref[table:correlation]{Table~\ref{table:correlation}}. If we exclude the interval IV, then we find a positive correlation with a Pearson $r^2$ coefficient of $\sim$0.8, with a $p$-value of 0.416, which is high as it arises from a small population. 
However, if interval IV is taken into account then the Pearson correlation coefficient is $\sim$0.52. 
In both cases we find a relationship with a positive $r^2$. Similarly, \citet{2019arXiv190602917B} reported a weak relationship between these two variables.

%\begin{table}[h!]
%\centering
%\resizebox{.5\textwidth}{!}{%
%\begin{tabular}{@{}lcccc@{}}
%\toprule
%Interval & I       & II      & III     & IV      \\ \midrule
%Energy & $38.2\pm4.5$ & $36.7\pm3.5$ & $28_{-27}^{+17}$ & $31.3\pm2.6$ \\ 
%Luminosity & $8.76\pm0.3$ & $6.01\pm0.25$ & $5.03\pm0.13$ & $8.23\pm0.3$ \\ \bottomrule
%\end{tabular}
%}
%\caption{Estimated cyclotron line energies from spectral fits in units of keV. %Luminosity is in units of $10^{35}~{\rm erg\,s}^{-1}$ with an assumed distance of $2.2_{-0.7}^{+0.5}$~kpc. A positive relationship was determined between both quantities.}
%\label{table:correlation}
%\end{table}

\begin{table}[h!]
\centering
\resizebox{.5\textwidth}{!}{%
\begin{tabular}{@{}lcccc@{}}
\toprule
Interval & I       & II      & III     & IV      \\ \midrule
Energy & $38.2\pm4.5$ & $36.7\pm3.5$ & $28_{-27}^{+17}$ & $31.3\pm2.6$ \\ 
Luminosity & $8.76\pm0.3$ & $6.01\pm0.25$ & $5.03\pm0.13$ & $8.23\pm0.3$ \\ \bottomrule
\end{tabular}
}
\caption{Estimated cyclotron line energies from spectral fits in units of keV. Luminosity is in units of $10^{35}~{\rm erg\,s}^{-1}$ with an assumed distance of $2.2_{-0.7}^{+0.5}$~kpc. A positive relationship was determined between both quantities.}
\label{table:correlation}
\end{table}

\raggedbottom

\subsection{\oao in the Corbet Diagram} \label{sec:corbet}

The high mass X--ray pulsar binaries (HMXBs) can be classified as Roche lobe-filling supergiants, wind accretion supergiant and Be-HMXB. A large number of HMXBs are mostly Be X--ray binaries \citep{2006A&A...455.1165L, 2013AdSpR..52.2132C}.

The Corbet diagram compares the orbital period to the spin period of the accreting pulsar \citep{1984A&A...141...91C, 1986MNRAS.220.1047C}. 
This can be interpreted as a "tidal" lock between the rotational velocity of the magnetospheric radius and the keplerian velocity, i.e. the equilibrium period. 
For short  spin periods, matter cannot be accretted due to the propeller mechanism \citep{1975A&A....39..185I}. For spin periods longer than the equilibrium period, matter can be accreted into the pulsar, thus reducing the angular momentum and hence the spin period.

\begin{figure}[h!] \centering
    \includegraphics[width=\columnwidth]{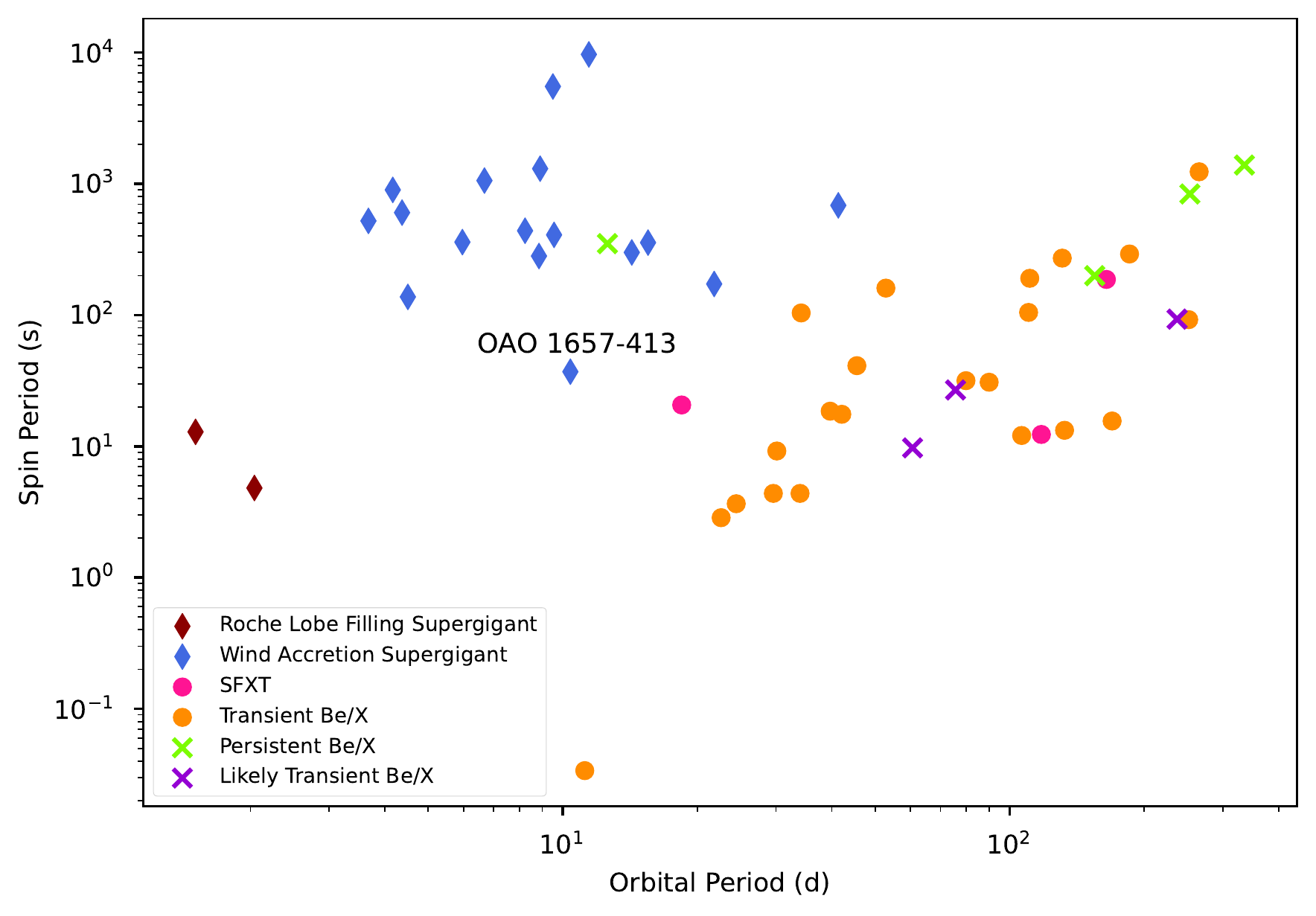}
    \caption{Corbet diagram for the different population of NS-HMXB with measured $P_{orb}$ and $P_{spin}$ periods. 
    Brown diamonds represent the roche lobe filling supergiant systems. Blue diamonds represent the wind accretion supergiant systems. Pink dots represent the SFXTs. Orange dots represent the transients Be/X systems while green crosses represents the persistent Be/X. Violet cross represents the transient Be/X candidates. Diagram adapted from \citealt{2012ApJ...759..124J}.}
    \label{Fig:corbet}
\end{figure}

The values of the spin and orbital periods of \oao puts it in an interesting position in the Corbet diagram; in between three types of sources: wind accretion systems, Roche lobe overflow systems and Be X--ray binaries \citep{1993ApJ...403L..33C} (see \hyperref[Fig:corbet]{Figure~\ref{Fig:corbet}}). 
The position in the Corbet diagram is related to ITS spectral type: Ofpe/WNL. These stars exhibit slower wind velocities and higher mass loss rates \citep{2007HiA....14..207M}. The combination of these characteristics allows for a higher accretion rate, and thus transferring angular momentum to the pulsar \citep{2009A&A...505..281M}. This spectral type occurs at the transition of main sequence OB stars and hydrogen-depleted Wolf-Rayet stars.  \cite{1994PASP..106.1025H} proposed that this spectral type is the hot quiescent state of the luminous blue variables stars (LBVs). This could explain the significant X-ray luminosity variability over months, as it would increase the mass loss rate or stellar radius, generating a closer approach to the Roche lobe and therefore increasing the mass transfer rate (see \cite{2007A&A...466..595K})

%-------------------------------------------------------------------

\section{Conclusions}
\label{sec:concl}

%In this work we study an observation of \oao observed by \textsl{NuSTAR} and \textsl{Swift}/BAT in June 2019, during the brightest orbital phase. 
In this work we analyzed the spectral and timing properties of \oao observed by the {\it NuSTAR} observatory on June 10, 2019, during the brightest orbital phase. We also included the {\it Swift}/BAT cumulative light curve spanning $\sim$16~years.
The {\it NuSTAR} light curve was divided into four intervals in order to characterise each spectral period. Power spectra were extracted for each interval, detecting the NS spin period (pulse) in all of them. At the same time, different light curves were taken in different energy ranges in order to analyse the extent to which the pulse was detected. We find that the pulse is detected up to $\sim$~80 keV in all time intervals.
% We find that the pulse increased throughout the whole observation by 0.011s which can be explained as the Doppler shifting by the orbital motion of the neutron star. 

The spectrum of \oao can be approximately described by a power law with an exponential cutoff. The comptonisation of soft X-rays above the surface of the neutron star can explain the continuum emission. Iron K$\alpha$ and K$\beta$ emission lines are present on the spectra. No emission lines associated with ionised material were detected \citep{2014MNRAS.442.2691P}, probably due to the low/moderate spectral resolution of {\it NuSTAR}. 
At the same time, a flux deficit at around 35~keV was observed in the residual spectrum. This indicates the presence of a cyclotron absorption feature. An absorption model was included, resulting in a better fit than the continuum and emission lines only.  We found that the cyclotron absorption line has a significance of $\sim$~3.4$\sigma$. This absorption line coincides with one found by \citet{1999A&A...349L...9O} detected with {\sl Beppo-SAX}. We interpret this second finding as evidence of the presence of this absorption feature on the x-ray spectra of \oao. 

By obtaining the energy centroid of the absorption line, the magnetic field on the surface of the neutron star was estimated to be $4.0\pm0.2\,\times\,10^{12}$~G. 
By combining the magnetic field and the x-ray flux,  we get an estimated distance lower limit of $\gtrsim$1~kpc. This distance is compatible with the distance obtained from {\sl Gaia} observations, and in turn from near-infrared observations estimated by \citet{2009A&A...505..281M}.
%The method we used for distance calculation is flux dependent, therefore less accurate and thus giving larger error bars that the ones previously mentioned.
%The energy associated with the cyclotron absorption line can change as a function of luminosity, we studied whether there was a possible correlation, either positive or negative. 
As already seen in other sources, we find a possible positive relationship between the cyclotron energy and the X--ray luminosity. 
Finally, we comment on the striking position of \oao in the Corbet diagram. This position can be attributed to its spectral type, which also explains the significant variability in its X-ray luminosity.

\begin{acknowledgements}
      
We thank the anonymous reviewer for their valuable comments on this manuscript. FAF, JAC and FG acknowledge support by PIP 0113 (CONICET). FAF is fellow of CONICET. JAC and FG are CONICET researchers. This work received financial support from PICT-2017-2865 (ANPCyT). This work was partly supported by the Centre National d'Etudes Spatiales (CNES), and based on observations obtained with MINE: the Multi-wavelength INTEGRAL NEtwork. JAC was also supported by grant PID2019-105510GB-C32/AEI/10.13039/501100011033 from the Agencia Estatal de Investigaci\'on of the Spanish Ministerio de Ciencia, Innovaci\'on y Universidades, and by Consejer\'{\i}a de Econom\'{\i}a, Innovaci\'on, Ciencia y Empleo of Junta de Andaluc\'{\i}a as research group FQM-322, as well as FEDER funds.

\end{acknowledgements}

% WARNING
%-------------------------------------------------------------------
% Please note that we have included the references to the file aa.dem in
% order to compile it, but we ask you to:
%
% - use BibTeX with the regular commands:
%   \bibliographystyle{aa} % style aa.bst
%   \bibliography{Yourfile} % your references Yourfile.bib
%
% - join the .bib files when you upload your source files
%-------------------------------------------------------------------
\bibliographystyle{aa}
\bibliography{aanda}

\end{document}